\documentstyle[12pt]{article}
\begin{document}
\topskip 2cm 
\begin{titlepage}

\begin{flushright}
Liverpool Preprint: LTH 370\\
 hep-lat/yymmddd\\
 7 May 1996\\
 \end{flushright}
 
\begin{center}
{\large\bf GLUEBALLS AND HYBRID MESONS} \\
\vspace{2.5cm}
{\large Chris Michael  
\footnote{
presented at the Rencontre de Physique `Results and Perspectives in 
Particle Physics',   
La Thuile,  March 4 1996}
		      } \\
\vspace{.5cm}
{\sl Theoretical Physics, Dept. of Mathematical Sciences, University of
Liverpool, }\\
{\sl Liverpool L69 3BX, UK }\\
\vspace{2.5cm}
\vfil
\begin{abstract}

The status of non-perturbative QCD calculations for mesons with gluonic 
excitation is presented. Lattice results for  the glueball spectrum are
reviewed. For hybrid mesons, the heavy quark  results are summarised and
new results are presented for light  quarks. Preliminary results  for
the spectrum of light-quark hybrid mesons indicate substantial  mixing
with quark model states for non-exotic $J^{PC}$. For the exotic 
$J^{PC}$ hybrid mesons,  the  $J^{PC}=1^{-+}$, $0^{+-}$ and $2^{+-}$
states are explored.

\end{abstract}

\end{center}
\end{titlepage}

\section{Introduction}

Since the advent of QCD as a theory of hadronic interactions, there have
been experimental searches for unambiguous evidence of gluonic
excitations in mesons. These searches need to be guided by theoretical
input.  The theoretical exploration involves non-perturbative methods
and lattice  QCD has become the most reliable tool. Here we review the
status of  glueball mass determinations from the lattice. The main
aspect of topicality  comes from a widely publicised claim~\cite{wein2}
that the lattice work uniquely targets a particular experimental 
candidate. We discuss this claim and put it in context.

Another area which is promising for a study of gluonic excitations  is
that of hybrid mesons. These have a gluonic field in a non-trivial 
representation so that it is truly excited. We review lattice results 
for this spectrum for the case of heavy quarks. New results for light 
quarks are presented. These preliminary results give strong evidence 
for the splitting among the many possible hybrid meson states.  The 
states with exotic quantum numbers (ie not allowed by  the naive quark
model:   $J^{PC}=1^{-+}$, $0^{+-}$ and $2^{+-}$) are studied and 
their spectrum is estimated.

\section{Glueball Masses}

The difficulty in isolating glueball candidates experimentally comes 
from the indirect methods that have to be used to deduce if a  given
resonance is composed primarily of gluons or of quarks. Lattice  QCD
allows the quark masses to be varied at will. In the simplest case,  the
quenched approximation, the dynamical quark mass is taken as large so 
that no quark loops are present in the vacuum.  In this approximation,
glueballs  are stable and do not mix with quark - antiquark mesons. 
This approximation  is very easy to implement in lattice studies: the
full gluonic action  is used but no quark terms are included. This
corresponds to a full  non-perturbative treatment of the gluonic degrees
of freedom in  the vacuum. A systematic lattice study of the neglected
quark loop effects can  be made in principle - though no comprehensive 
treatment has yet been made.

The glueball mass can be measured on a lattice through evaluating the
correlation $C(t)$ of two closed  colour loops (called Wilson loops) at
separation $t$ lattice  spacings.   This correlation has contributions 
from all glueballs of the given symmetry, with the ground state
contribution  dominating at large $t$. In practice, sophisticated
methods are used to choose loops  such  that the correlation $C(t)$ is
dominated by the ground state glueball. By using  several different
loops,  a variational method can be used to achieve this effectively.  
Even so, it is worth keeping in mind that upper  limits on the ground
state mass are obtained in principle. 

The method also needs to be tuned to take account of the many 
glueballs: with different $J^{PC}$ and different momenta. On the lattice
the Lorentz symmetry is reduced to that of  a hypercube. Non-zero 
momentum sates can be created (momentum is discrete in units of 
$2 \pi /L$ where $L$ is the lattice spatial size). The usual 
relationship between energy and momentum is found for sufficiently 
small lattice spacing. Here we shall concentrate on the simplest 
case of zero momentum (obtained by summing the correlations over 
the whole spatial volume).

For a state at rest, the rotational symmetry becomes a  cubic symmetry.
The lattice states will transform  under irreducible
representations of this cubic symmetry group (called $O_h$). These
irreducible representations can be linked to  the representations of the
full rotation group SU(2). Thus, for  example, the five spin components
of a $J^{PC}=2^{++}$ state should be appear as  the two-dimensional 
E$^{++}$ and the three-dimensional T$_2^{++}$ representations on  the
lattice, with degenerate masses. This degeneracy requirement then
provides a  test for the restoration of rotational invariance -  which
is expected to occur at sufficiently small lattice spacing.

\begin{figure}[p]
\vspace{14cm} 
\includegraphics{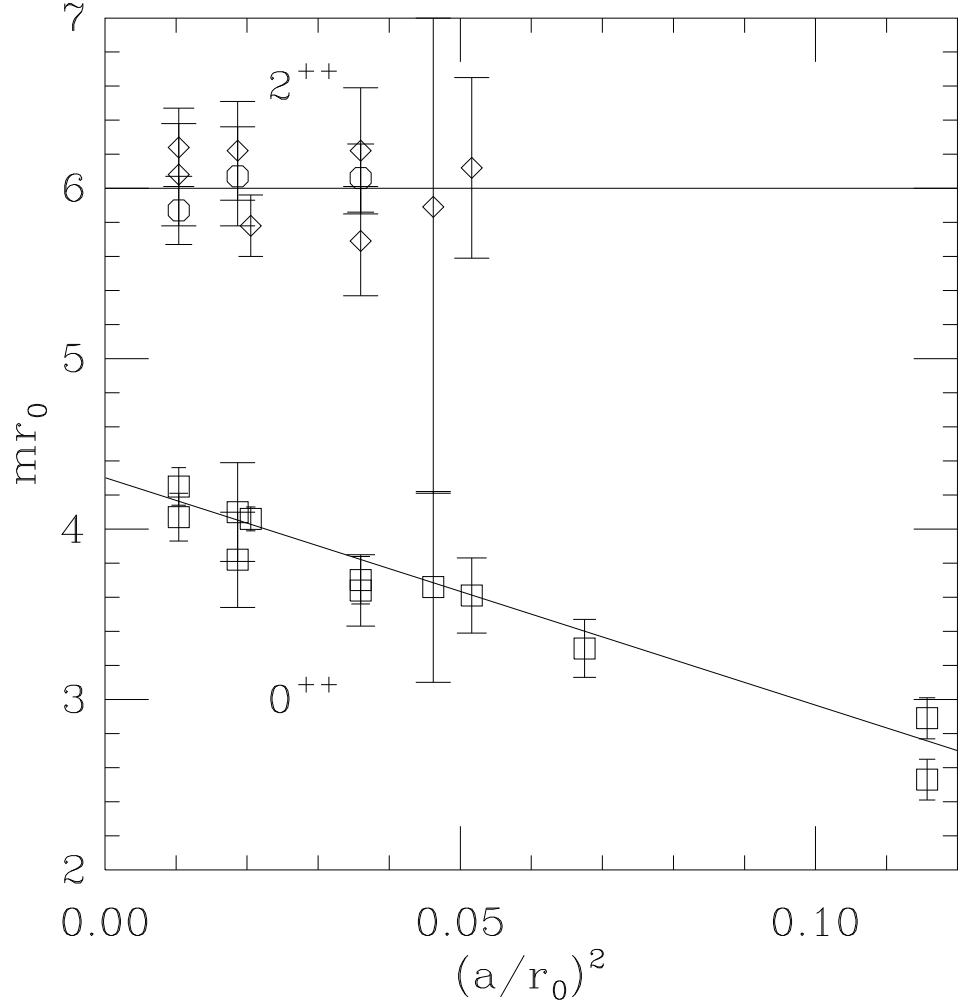}
 \caption{ The value of mass of the  $J^{PC}=0^{++}$ and $2^{++}$
glueball states from refs{\protect\cite{DForc,MT,glue,gf11}} in units of
$r_0$. The restoration of rotational  invariance is shown by the
degeneracy of the $T_2$ and $E$  representations that make up the
$2^{++}$ state: shown  by  octagons  and diamonds respectively.  The
straight lines  show  fits describing the approach to the continuum
limit as $a \to 0$.
   }
\end{figure}

The results of lattice measurements~\cite{DForc,MT,glue,gf11} of the
$0^{++}$ and  $2^{++}$ states are shown in fig~1.  Since the lattice
observables,  such as the glueball mass $\hat{m}$, are not in physical
units, it is necessary  to form dimensionless ratios of lattice
observables to compare with  experiment.  Fig~1 shows the dimensionless
combination of the  lattice glueball mass $\hat{m}_0$ with a lattice
quantity $\hat{r}_0$, which is a well measured quantity (given  by
$\hat{r}^2 d\hat{V}/d\hat{r} = 1.65$ at $\hat{r}=\hat{r}_0$ where
$\hat{V}(\hat{r})$ is the lattice interquark potential at separation
$\hat{r}$) on the lattice that can be used to calibrate the lattice
spacing and so explore the continuum limit. The quantity plotted,
$\hat{m}_0 \hat{r}_0$, is expected to be equal to the product of
continuum quantities $m_0 r_0$ up to corrections of order $a^2$. This 
behaviour near the continuum limit is indeed found as shown  by the
linear dependence of fig~1. The extrapolation to the continuum limit ($a
\to 0$) can now be made with confidence. Note that older lattice data 
were only available at larger values of $a^2$ which explains why a
smaller $0^{++}$  glueball mass was favoured at that time.

The lattice results in fig~1 from the UKQCD and GF11 groups  have
signals which are  of comparable statistical significance and which are
consistent with each other. However, their published
values~\cite{glue,gf11} of the $0^{++}$ glueball mass are different
(1550 versus 1740 MeV). The GF11 group chose to extrapolate
$\hat{m_0}/\Lambda a$ to the continuum limit. This ratio has  the
disadvantage that  there can be corrections both of order $a$ and of
order $(\ln a)^n$ while they  assume in their extrapolation that only
order $a^2$ effects are significant. They determine $\Lambda a$ from
their own results for $\hat{m}_{\rho}$ which yields a  $0^{++}$ mass of
$1740 \pm 71$ MeV leading them to  claim~\cite{gf11,wein2} that the $f_J
(1710)$ meson is a preferred glueball candidate.  Their error  estimate
on the glueball mass  does not take into account fully the systematic
errors in the  extraction of the continuum limit or those due to
quenching. 

Using instead the best determined {\it continuum} quantity from the 
lattice results, we need to determine a physical value for $r_0$. From the 
interquark potential as determined in spectroscopy, the value of $r_0$
in physical units is about 0.5 fm and we will  adopt a scale equivalent
to $r_0^{-1} = 0.372 $ GeV.  This information yields lattice 
predictions for the glueball masses based on all lattice data of around
1.6 GeV and 2.2 GeV  for the $0^{++}$ and $2^{++}$ glueballs
respectively. Setting the scale in a quenched lattice calculation is
inherently imprecise  because ratios of lattice observables are found to
disagree with experimental values (unquenched) by  different amounts for
different ratios. Thus no common scale determination is possible for the
quenched lattice.   It is prudent to assign a systematic error of at
least 10\% to the scale. Since this dominates the statistical error, the
conservative conclusion  is  a  $0^{++}$ glueball mass of $1600 \pm 160$
MeV. This is an energy range consistent with promising experimental
$0^{++}$ glueball  candidates such as $f_0(1500)$ - for a review see
ref\cite{close}. A candidate for a $2^{++}$ glueball at 2230 MeV has
also been reported  recently~\cite{close}.

\begin{figure}[p]
\vspace{16cm}
\includegraphics{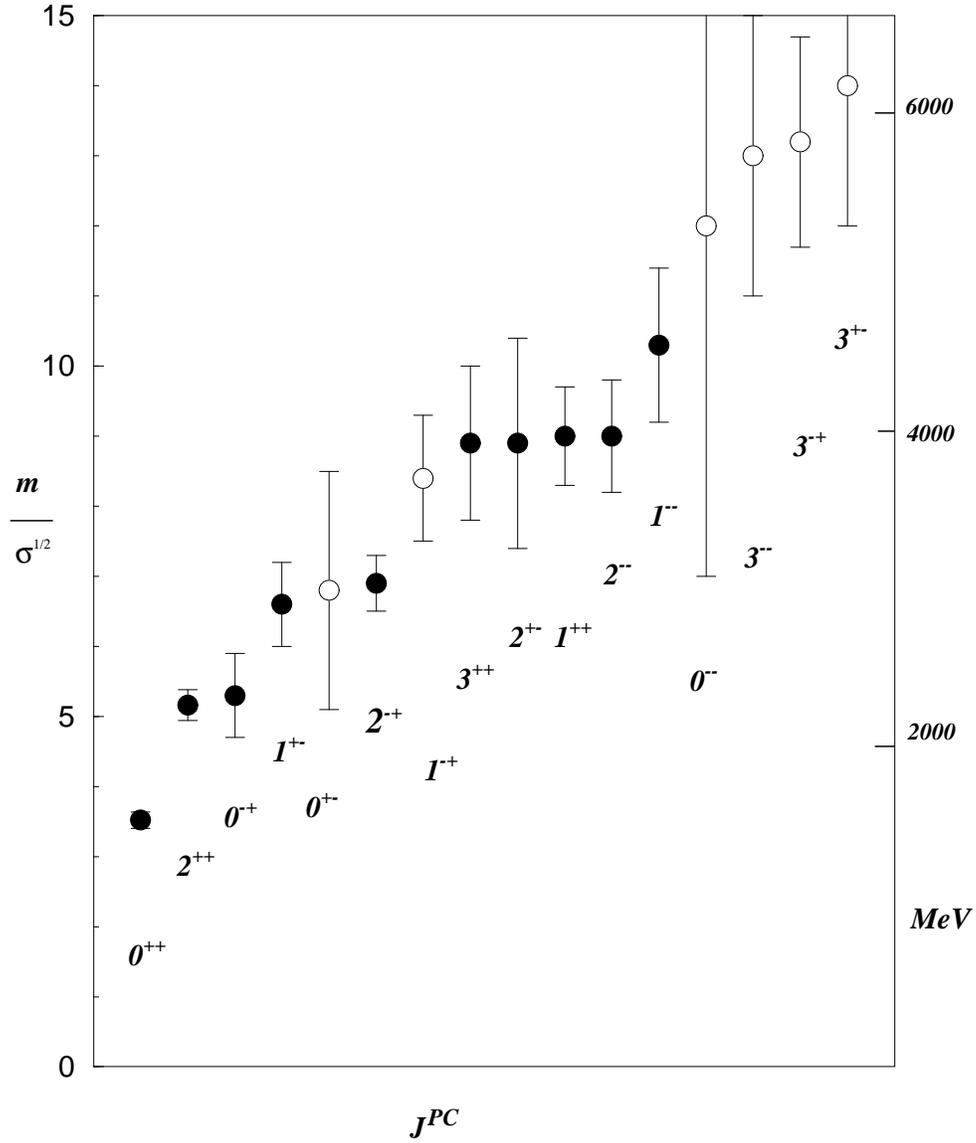}
  \caption{ The  mass of the  glueball states  with quantum  numbers
$J^{PC}$ from ref{\protect\cite{glue}}.  The scale is set by
${\protect\sqrt{\sigma}} \approx 0.44 $ GeV which yields the right hand
scale in GeV. The solid points represent mass determinations whereas the
open points  are upper limits.
   }
\end{figure}

The predictions for the other $J^{PC}$ states are that they lie  higher
in mass and the present state of knowledge is summarised  in fig~2. 
Note that the lattice gives a clear indication that no light 
pseudoscalar glueball should exist. Remember that the lattice results
are strictly  upper limits. For the   $J^{PC}$ values not shown, these 
upper limits are too weak to be of use. 

Since quark - antiquark mesons can only have certain  $J^{PC}$ values, 
it is of special interest to look for glueballs with  $J^{PC}$ values 
not allowed for such mesons: $0^{--}, 0^{+-}, 1^{-+}, 2^{+-}, $ etc. 
Such spin-exotic states, often called ``oddballs'',  would not mix
directly with quark - antiquark mesons. This would  make them a very
clear experimental signal of the underlying glue dynamics.  Various 
glueball models (bag models, flux tube models, QCD sum-rule  inspired
models,...) gave different predictions for the presence of such oddballs
(eg. $1^{-+}$) at relatively low masses. The lattice mass spectra
clarify these uncertainties but, unfortunately for experimentalists,  do
not indicate any low-lying oddball candidates. The lightest candidate 
is from the T$_2^{+-}$ spin combination. Such a state could correspond 
to an $2^{+-}$ oddball. Another interpretation is also possible,
however,  namely that a non-exotic $3^{+-}$ state is responsible (this
choice of interpretation can be resolved in principle by finding the 
degenerate 5 or 7 states of a $J=2$ or 3 meson). The overall  conclusion
at present is that there is no evidence for any oddballs  of mass less
than 3 GeV.

Glueballs are defined in the quenched approximation and, hence, they  do
not decay into mesons since that would require quark - antiquark 
creation. It is, nevertheless, still possible to estimate the  strength
of the matrix element between a glueball and a pair of mesons  within
the quenched approximation. For the glueball to be a relatively narrow 
state, this matrix element must  be  small. A  very preliminary
attempt~\cite{gdecay,wein2} has been made to estimate the size of  the
coupling of the $0^{++}$ glueball to two pseudoscalar mesons. A
relatively small value is found. Furthermore  they see indications for a
dependence on the pseudoscalar mass of the  reduced decay matrix
element.  These conclusions imply that the quenched glueball  mass
determination was of relevance to the experimental situation since  the
mixing with other mesons would be small.  Further work needs to be done
to  investigate this in more detail, in particular to study the mixing
between  the glueball and $0^{++}$ mesons since this mixing may be an
important  factor in the decay process.

In principle, it is possible to study on a lattice  
the glueball spectrum in full QCD vacua with sea quarks of mass $m_D$.
For large $m_D$, the result is just the quenched result described above. 
For $m_D$ equal to the experimental light quark masses, the results 
should just reproduce the experimental meson spectrum - with the 
resultant uncertainty between glueball interpretations and other 
interpretations. The lattice enables these uncertainties to be resolved
in principle: one obtains the spectrum for a range of values of $m_D$ 
between these limiting cases, so mapping glueball states at large 
$m_D$ to the experimental spectrum at light $m_D$.

\section{Hybrid Mesons}

\begin{figure}[p]
\vspace{14cm} 
\includegraphics{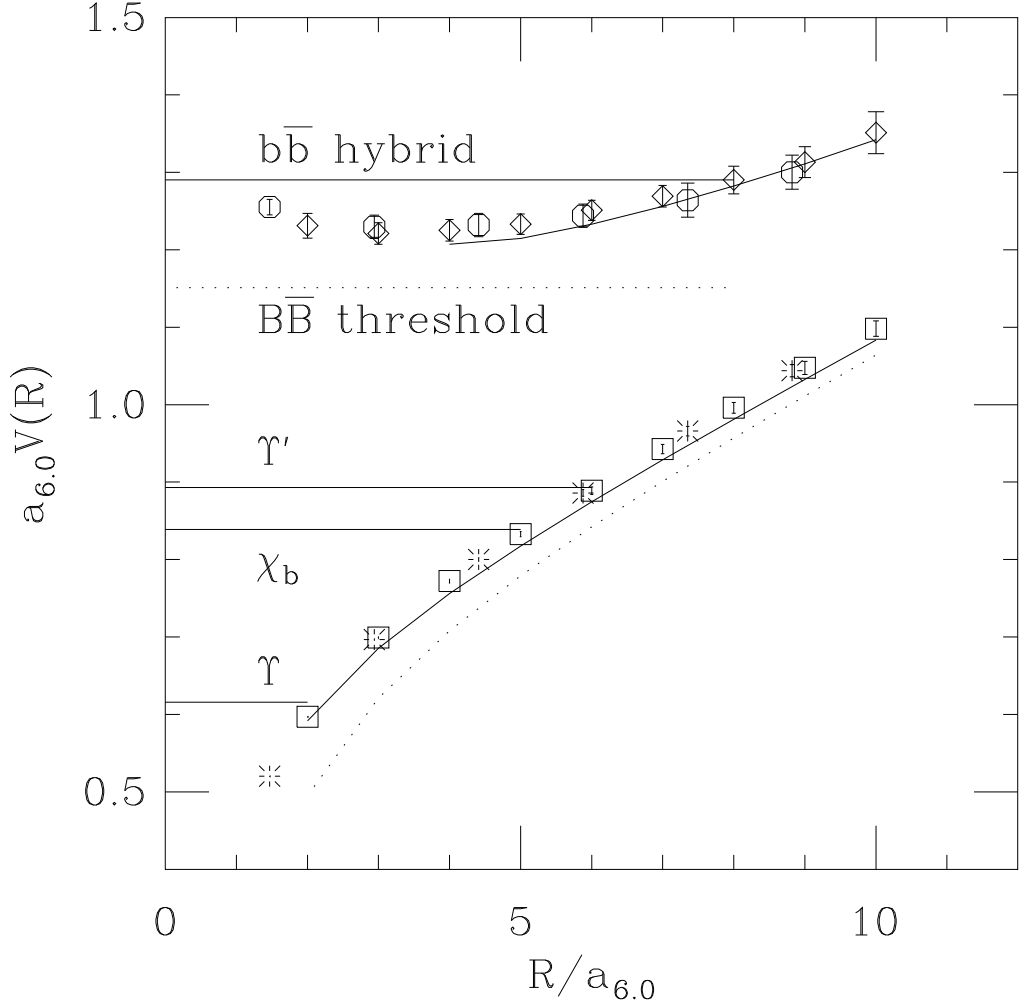}
 \caption{ The lattice static quark potential for the  ground state and
first excited state from ref~{\protect\cite{cmper}} with the scale 
given by a lattice spacing corresponding to $a_{6.0}=0.50 $ GeV$^{-1}$. 
The energy difference between  the excited potential and the ground
state is  seen to be well approximated~{\protect\cite{cmper}} by a
string model expression ($\approx \pi/R$ as shown by  the continuous
line). Also shown are some  of the lower lying $b\bar{b}$ states  in
these potentials obtained from  the Schr\"odinger  equation in the
adiabatic approximation. The lattice potentials share a common
self-energy so that the energy difference between the lowest hybrid
level and the $\Upsilon$ meson is determined directly (1.36 GeV). The
dotted curve shows the modification to the  quenched lattice
ground-state potential needed to give the experimental  spectrum. This
gives an estimate of the systematic error from  quenching.
   }
\end{figure}

 In order to set the scene for the study of mesons with  gluonic
excitations, it is worthwhile to summarise briefly the  simple
constituent quark model. This model of massive quark and  antiquark
bound by a potential is only justified theoretically  for $b\bar{b}$ 
and to a lesser extent $c\bar{c}$, but it is still useful guide for 
light quark states.  The mesonic states that can be  made from
$q\bar{q}$ with spatial wavefunction with orbital angular momentum $L$ 
and total spin $S^{PC}=0^{-+}$ or $1^{--}$ have $J^{PC}$ values of
 \smallskip

\halign{\hskip 0.5in # \hfil & \qquad $#$ \hfil \cr
$L^{PC}=0^{++}$ & J^{PC}=0^{-+}, 1^{--}   \cr
$L^{PC}=1^{--}$ & J^{PC}=1^{+-}, 0^{++}, 1^{++}, 2^{++}  \cr
$L^{PC}=2^{++}$ & J^{PC}=2^{-+}, 1^{--}, 2^{--}, 3^{--}. \cr
}
 
\smallskip
 \noindent Since the gluon can introduce no flavour quantum numbers, the
$J^{PC}$ assignments will be of importance. Of special interest in the
following will be the absence of certain $J^{PC}$ values in the above
list. These states are known  as spin-exotic and include  $0^{--},
0^{+-}, 1^{-+}, 2^{+-} $.

We define a hybrid meson as a $q\bar{q}$ system with additional  gluonic
excitation.  The definition of a  hybrid meson is less clear than for  a
glueball since even the basic  quark model mesons have a gluonic
component which is responsible  for the binding force. So we must
establish that the gluonic  component is excited before labelling a
state as a hybrid meson. This is straightforward for the  case of static
quarks at separation $R$.  The ground state potential  will then have
cylindrical symmetry about the interquark axis while less symmetric
configurations correspond to  various excitations of the gluonic flux
joining the sources. A pioneering study  using lattice 
techniques~\cite{liv} found that the first excited gluonic state arises 
from transverse gluonic flux excitations of the form  $\, \sqcap -
\sqcup$. Such  spatial excitations of the  distribution of the colour
flux from  quark to antiquark correspond to  gluonic fields with $J_z
\ge 1$ about the interquark axis and so are clearly hybrid states.

In molecular physics, it is common to assume that the electronic 
degrees of freedom adjust themselves with a much shorter timescale than 
that of the rotation of the molecule as a whole - this is the adiabatic 
approximation. For hybrid mesons, this will be valid  if the gluonic
degrees of freedom have a much shorter time-scale than those  associated
with the quarks. This will be a plausible approach since we find gluonic
excitations with  energies exceeding 1 GeV while quark model excitations
(orbital and radial)  have  smaller energies (of a few hundred MeV). 
Then the   allowed $J^{PC}$ values of  hybrid mesons bound in this
excited  gluonic potential can be easily determined within this
adiabatic approximation  using the Schr\"odinger equation.  The lowest
lying hybrid states are found~\cite{liv}  to have $J^{PC}$ values
arising from  two spatial symmetries
 \smallskip

\halign{\hskip 0.5in # \hfil & \qquad $#$ \hfil \cr
$L^{PC}=1^{+-}$ &  J^{PC}=1^{--}, 0^{-+}, 1^{-+}, 2^{-+}   \cr
$L^{PC}=1^{-+}$ &  J^{PC}=1^{++}, 0^{+-}, 1^{+-}, 2^{+-}   \cr
}
 
\smallskip
 \noindent The first group corresponds to  the states accessible from a
``magnetic gluon'' excitation with spatial symmetry  $L^{PC}=1^{+-}$
while the second group are from an ``electric gluon'' with
$L^{PC}=1^{-+}$. The lattice determination~\cite{cmper} of the  ground
state and excited potentials is illustrated in fig~3 for the quenched 
case. At large interquark separation $R$, a hadronic string picture  is
expected to be a reasonable model and we see that the string model 
excitation energy ($\pi/R$ in the simplest version - but see
ref~\cite{cmper}) gives  a good description.  Also  shown is the
spectrum of mesons in these potentials obtained in the abiabatic
approximation  for $b$ quarks.  Although the absolute value of the bound
state energy is  not  accessible because of lattice self-energy effects,
the difference  between energies of bound states in the ground and
excited potential  is completely predicted. The hybrid level  shown is
the lowest such level and has the above  eight degenerate $J^{PC}$
values. Thus there will be mesons with exotic quantum numbers  at this
energy which provide a prediction that can be checked by experiment.  

The ground state potential in the quenched approximation does not 
correctly reproduce the experimental $\Upsilon$ spectrum. The simplest 
explanation is that in full QCD the short distance (Coulombic) component
of the potential would be enhanced (by $33/(33-2N_f)$ at leading order
in perturbation theory) and such an enhancement  is illustrated in fig~3
by the dotted ground state potential that does indeed  reproduce the
experimental spectrum.  Using this prescription, but   taking into
account uncertainties from different approaches to modifying the
quenched  approximation, the lattice prediction~\cite{cmper} is for the
lightest hybrid meson  excitation to be at 4.19(15) GeV for $c \bar{c}$
and 10.81(25) GeV  for $b \bar{b}$.   These energy values lie above the
open $D \bar{D}$ and $B \bar{B}$ thresholds. An alternative  procedure,
within the quenched approximation, is to focus~\cite{sommerb} on the
energy  difference between the hybrid meson and the $B\bar{B}$ 
threshold. This suggests that the lowest hybrid level may lie below  the
threshold.

These lattice predictions do not take account of splitting of the 
degeneracy of the hybrid levels due to spin-spin and spin-orbit effects.
Indeed, going beyond the adiabatic approximation, the non-exotic hybrid
mesons can mix with states in the usual  $L$-excited quark model and may
 thus be modified substantially. Evidence also exists from lattice 
studies~\cite{gluelump} at small separation $R$ that the  
$L^{PC}=1^{+-}$ excitation  is a few hundred MeV lighter than the
$L^{PC}=1^{-+}$ excitation. This would imply that the degenerate hybrid
levels would  be split with the lightest exotic state having
$J^{PC}=1^{-+}$.   The experimental detection of such states depends
crucially on whether  they lie above or below the open quark threshold.
The quenched lattice estimate~\cite{cmper}  shows that, for both $b$ and
$c$ quarks, the lightest hybrid levels lie above threshold. The
uncertainties due to the level splitting  effects described above,
combined with the uncertainty in interpretation of the quenched
spectrum, both  point to the possibility that a narrow spin exotic
hybrid meson could exist close to the open  quark threshold. It  is
important to search carefully for such states.

\begin{figure}[p]
\vspace{14cm} 
\includegraphics{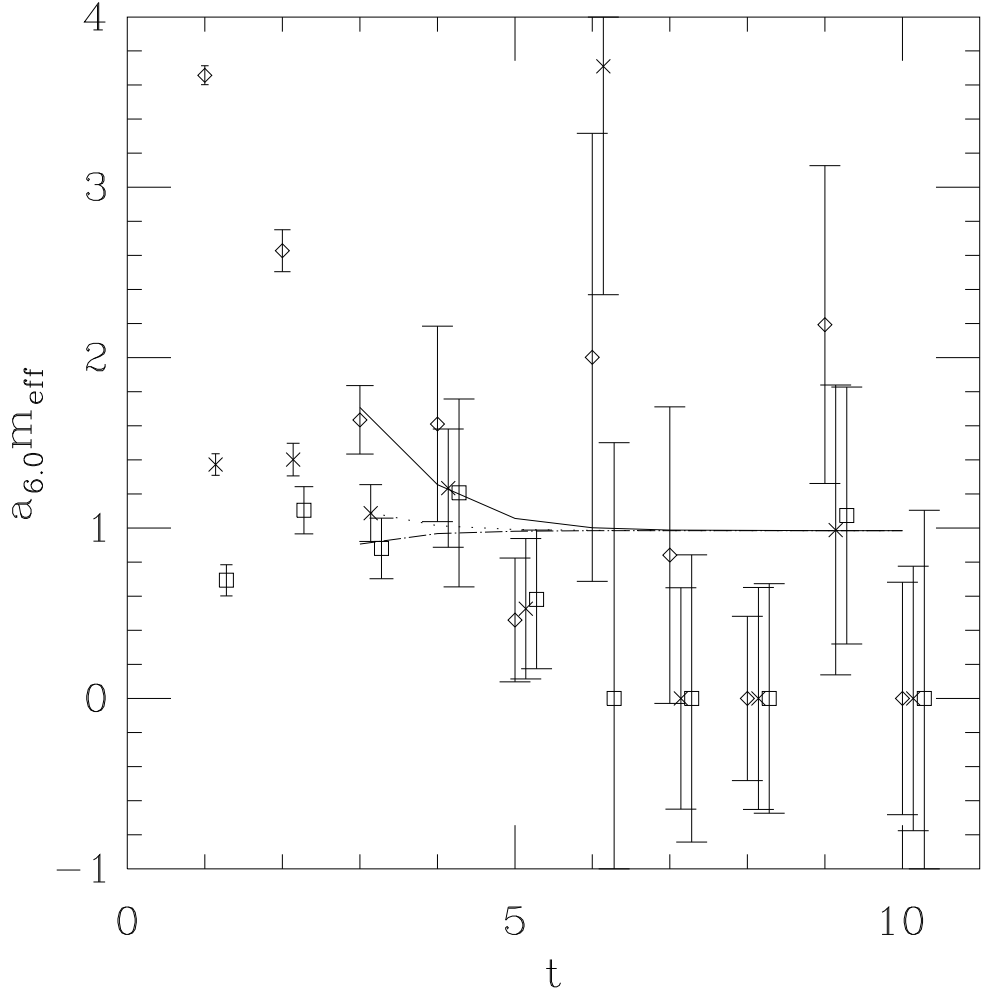}
 \caption{ The lattice effective mass for the $J^{PC}=1^{-+}$  hybrid
meson versus time separation $t$. The source used was a U-shaped path of
size  $6 \times 6$ while the sinks were combinations of U-shaped paths 
of size $6 \times 6$ (diamonds), $3 \times 3$ (crosses) and $1 \times 1$
(squares). The fit shown has a ground state mass  of 0.98(26) in lattice
units at $\beta=6.0$ with tadpole-improved  clover action for hopping
parameter $K=0.137$. This corresponds to  mesons made of strange quarks
so $m(1^{-+})/m(\phi)=1.8(5)$. 
   }
\end{figure}

\begin{figure}[p]
\vspace{14cm} 
\includegraphics{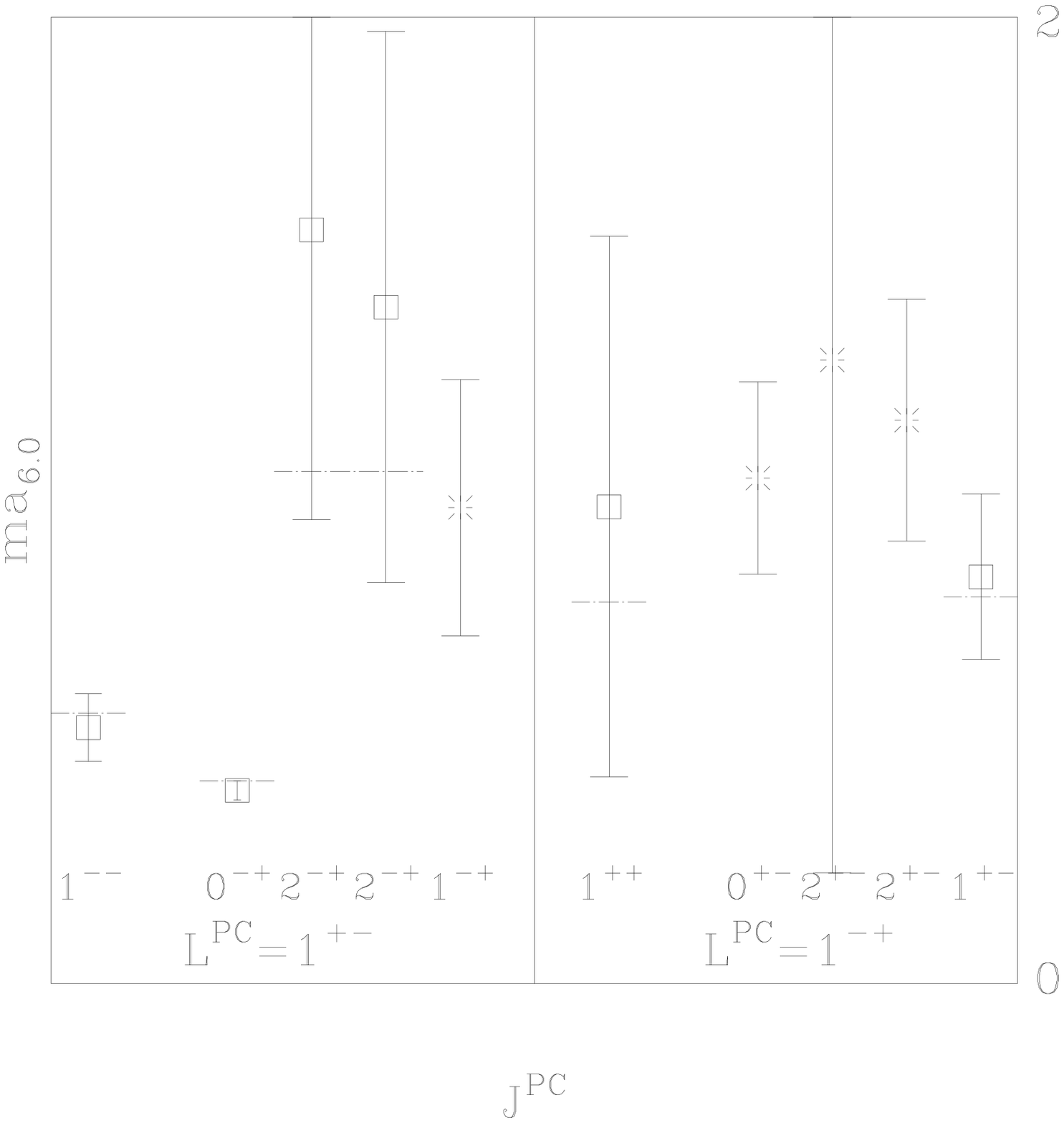}
 \caption{ Preliminary results for the ordering of the hybrid  meson
levels for strange quarks~{\protect\cite{lm}}. The states with  burst
symbols are $J^{PC}$ exotic. The dashed lines represent $L$-excited
quark model states as determined on the lattice. The strong mixing of
the states  created by our hybrid operators  with these is apparent for
the pseuodoscalar and vector meson  cases. 
   }
\end{figure}

In principle the lattice approach allows a study of hybrid mesons 
formed from light ($u$, $d$ and $s$) quarks. Preliminary results have 
recently been obtained~\cite{lm} by the UKQCD collaboration. The method
builds on the experience with static quarks and uses  operators to
create hybrid mesons in which the quark and antiquark  are joined by
colour flux which is excited in the transverse plane as  $\, \sqcap -
\sqcup$. Excitations of this kind are clearly  non-trivial gluonic
contributions and the mesonic states in  such excited potentials include
exotic $J^{PC}$ values.  The lattice analysis is a fully relativistic 
analysis of propagating mesons. The approximations used are that of the 
quenched approximation for the vacuum and, in this preliminary study, we
used a light quark mass corresponding to  the strange quark (since 
lighter quarks are computationally more demanding).  From the results 
we will be able to explore the splitting among hybrid meson levels 
as well as the mixing between non-exotic hybrid mesons and $q \bar{q}$ 
mesons.

Preliminary results~\cite{lm} come  from 70 lattices of size $16^3
\times 48$ at $\beta=6.0$ which is  only a small fraction of the
eventual statistical sample. We used a SW-clover fermionic action  with
clover coefficient $c=1.4785$ and hopping parameter $K=0.1370$. To
establish our methods, we  have studied the $L$-excited quark model
mesons for S, P and D waves  and successfully determined their energy
levels. The signal for the hybrid mesons  is weaker and our present data
sample  does  not  give precise estimates of the hybrid meson masses.
For example, our results  for  the $J^{PC}=1^{-+}$ meson are shown by
the fit in fig~4  and are consistent  with a mass ratio $m/m_{\phi} =
1.8(5)$ which shows the large  errors remaining.   What  is somewhat
better determined, however, are the  splitting effects.  We see in fig~5
significant mixing  of non-exotic hybrids created using our hybrid
operators with $q \bar{q}$ mesons of the same quantum numbers.  We
intend to explore this more fully in future.  We find  that the exotic
hybrids are all at comparable masses with the state with spatial
excitation   $L^{PC}=1^{+-}$  slightly lower in mass so that the
lightest exotic hybrid meson would have $J^{PC}=1^{-+}$.

The preliminary  study presented here was conducted using $s$-quarks.
Quenched  lattice studies of the QCD spectrum suggest that quark mass
effects  in the meson mass (or mass squared) are well described by a 
term linear in the quark mass.  Experimental meson masses are 
consistent with the ansatz that $M^2_{s\bar{s}} - M^2_{q\bar{q}} 
\approx 0.5 $ GeV$^2$ where $q$ means $u$ or $d$. This suggests that the
$u$, $d$  mesons will be around 160~MeV lighter than the $s \bar{s}$
mesons for masses around 1.5 GeV. Note that these lattice studies are 
for mesons made from two quarks of equal mass which are thus 
eigenstates of $C$. For unequal masses (eg. strange mesons) the lack  of
$C$ invariance masks the identification of spin exotic states.

There have been several experimental claims for  hybrid   mesons - for
reviews see refs\cite{close,chung}.  Our results suggest that  non
spin-exotic candidates may need re-appraisal since big mixing effects 
are possible. For the exotic  mesons, the favoured candidate to lie
lowest will  have $J^{PC}=1^{-+}$ and several experimental hints of such
states have  been reported.

\section{Conclusions}

We have summarised quenched lattice predictions for glueballs and hybrid
mesons. Recent developments include an estimate of glueball decay widths
and a first study of light quark hybrid mesons.  The study of the  light
quark hybrid mesons with considerable greater statistics is currently
under way.  The qualitative features from such predictions are an
essential  guide to the experimental exploration of such mesons. Lattice
studies  with dynamical quarks will enable better control of the
systematic  error from quenching - this is also in progress.

\section{Acknowledgements}
   
     I acknowledge the contributions made by   my colleagues in the
UKQCD collaboration, especially  Pierre Lacock, to the study of hybrid
mesons with light quarks.

\end{document}